\documentclass[twocolumn]{revtex4-1}
\usepackage{indentfirst}
\usepackage{graphicx}
\usepackage{amsmath}
\usepackage{amsfonts}
\usepackage{color}
\usepackage[sc]{mathpazo}
\usepackage{booktabs}
\usepackage{amssymb}
\usepackage{rotating}

\begin{document}

\title{Classical memoryless noise-induced maximally discordant mixed separable steady states}
\author{Ferdi Altintas}\email{ferdialtintas@ibu.edu.tr} \author{Arzu Kurt} \author{Resul Eryigit}\email{resul@ibu.edu.tr}  
\affiliation{Department of Physics, Abant Izzet Baysal University, Bolu, 14280, Turkey.}
\begin{abstract}
We have investigated the dynamics of quantum discord and entanglement for two qubits subject to independent global transverse and/or longitudinal memoryless noisy classical fields. Global transverse and/or longitudinal random fields are found
to drive the system to maximally discordant mixed separable steady states for suitable initial conditions. Moreover, two independent noises in the system are found to enhance both the steady state randomness and quantum discord in the absence of entanglement for some initial states.
\end{abstract}

\maketitle
\section{Introduction}
Entanglement has been considered as the most useful type of quantum correlation for 
quantum informational and computational tasks, until recently. Its role as a resource to enhance 
quantum versions of operations such as quantum teleportation, key distribution and etc., has been widely examined~\cite{tele}. 
Entanglement as a resource suffers from the decoherence problem; it degrades under the effect of environmental 
noise which is unavoidable for most of the practical cases. Although there are studied protocols to overcome 
such effects, new type of "quantumness" resources beyond entanglement for quantum information tasks are desirable~\cite{qdproto,dqc1,grover,rsp}. Several 
measures have been proposed to characterize and quantify quantum correlations beyond entanglement~\cite{howz,dvb,modi}. 
One such measure of quantum correlations, that is shown to be useful in DQC1~(deterministic quantum computation with one quantum bit)~\cite{dqc1}, Grover search~\cite{grover} algorithms and remote state preparation~\cite{rsp}, was introduced by Ollivier and Zurek and is called quantum discord~(QD)~\cite{howz}. 
It is introduced as a mismatch of quantum version of two classically equivalent expressions of mutual information and has attracted a lot of attention, recently~\cite{facca,gessner,cabmpw,oisqd,expd,isd,qddeath,nfa,nfansc,lno,lnonsc}. In fact, it was shown by Ferraro {\it et al.,} that almost all quantum states of a bipartite system have non-zero quantum discord~\cite{facca} 
and studies on dynamics of QD in open quantum system indicate that QD is much more robust compared to entanglement under dissipative and decay processes. 
Although, the usage of quantum discord as a fundamental resource for quantum information processing and what it implies for the 
quantumness of a state are still under debate~\cite{gessner}, recently, quantum discord was given an operational interpretation in terms of entanglement consumption in an extended state merging protocol and dense coding contexts~\cite{cabmpw}.

Beside the characterization and quantification of the quantumness of correlations, the behavior of correlations under decoherence 
and/or dissipation have been also an active area of research~\cite{oisqd,expd,isd,qddeath,nfa,nfansc,lno,lnonsc}. Since any realistic
quantum system inevitably interacts with its environment, the environment-induced-noise may cause the system to lose  energy~(dissipation) and/or coherence~(dephasing).
It is expected that the quantum correlations would be easily destroyed by such unavoidable noises. In fact, entanglement is very fragile and can 
cease to exist in a finite time when the state of the system exceeds a certain level of mixture. This phenomena is known as entanglement sudden 
death~(ESD)~\cite{esd}. Contrary to ESD, the quantum correlations beyond entanglement, such as QD, are more robust against noise; they decay 
exponentially~\cite{expd} or suffer only instant death~\cite{isd}. Since the set of states with only classical correlations contain a tiny volume 
in the whole Hilbert space compared to that of the volume of the set of separable states~\cite{facca}, this, indeed, explains the robustness of QD 
to sudden death~(also see Ref.~\cite{qddeath} where it was demonstrated that for certain conditions QD can also suffer death for a finite time). 
Moreover, it was shown by many groups, both experimentally and theoretically, that QD exhibits noise-free-area for a finite time interval contrary 
to entanglement which dies very quickly~\cite{nfa}. Some necessary and sufficient conditions for constant (geometric) quantum discord for certain 
states under local noise have been given~\cite{nfansc}. Moreover, decoherence free subspaces~(DFS) also provide a mechanism to preserve quantum correlations perfectly and are investigated as another avenue for error-free quantum computation~\cite{dmmo}. In Ref.~\cite{lno}, local memoryless noise acting on only one party of the two qubit system 
initially in a fully classical state was found to create a significantly high QD, while entanglement cannot be created by such local and memoryless noise. 
Similarly, necessary and sufficient conditions for the local noise induced QD are discussed in Ref.~\cite{lnonsc}. 

Since entanglement is very fragile under the detrimental environmental effects, QD is hoped to be used as a fundamental resource for the 
noisy-implementation of quantum information protocols that rely on quantum correlations beyond entanglement~\cite{qdproto,dqc1,grover,rsp}. 
Therefore, finding maximally discordant mixed states that are robust against state mixedness is very desirable~\cite{aqdj,ggz,gpa}. 
For that purpose, recently, Al-Qasimi and James~\cite{aqdj} and Galve {\it et al.,}~\cite{ggz} have investigated the family of states 
for which QD can take a maximal value for a given entanglement. It was shown that the maximum value that QD can reach, for example, 
for separable states is 1/3. It is natural to ask: does the noise maximize the QD in the absence of entanglement ? To answer this question, 
in the present study, we consider two qubits subject to independent global transverse and longitudinal magnetic fields with noisy components 
in their amplitudes obeying white noise~(Markovian) assumption. We have shown that more independent noises in the system, although being both classical and
memoryless, can enhance, even maximize the value of the steady state quantum discord for separable states. 

The Letter is organized as follow. In Sec.~\ref{sec2}, by using cumulant expansion technique and Gauss-Markov approximation, we derive 
the master equation for two atoms subject to time-independent longitudinal and time-dependent transverse magnetic fields with noisy 
components in their amplitudes. The correlation measures, QD and entanglement, are also briefly discussed in this section. 
The dynamics of entanglement and quantum discord between two atoms initially in different product and Bell states under transverse 
and/or longitudinal noises is investigated in Sec.~\ref{sec3}. In Sec.~\ref{sec4}, we determine the initial states in which steady 
state QD is maximized for separable states by the one- or two-active-noise cases. We summarize the important results in Sec.~\ref{sec5}
\section{The Model and Correlation Measures}\label{sec2}
Here, we consider a system of two qubits placed in a magnetic field in $z$-direction and is also driven by a 
time-dependent sinusoidal transverse field. The semi-classical Hamiltonian can be given as~\cite{model1,model2}:  
\begin{eqnarray}\label{hamiltonian}
H(t)&=&\frac{1}{2}[\omega_{AB}(\sigma_z^A+\sigma_z^B)+\Omega\cos(\omega t)(\sigma_x^A+\sigma_x^B)\nonumber\\
&&+\Omega\sin(\omega t)(\sigma_y^A+\sigma_y^B)],
\end{eqnarray}
where $\omega_{AB}$ and $\omega$ are the atomic transition and transverse field frequencies, respectively, $\Omega$ is the Rabi frequency 
and $\sigma_{x,y,z}^{A,B}$ are the Pauli spin matrices. Here $\omega_{AB}$, $\omega$ and $\Omega$ are the system-control parameters 
and we assume that both $\omega_{AB}$ and $\Omega$ have a constant plus a randomly fluctuating part as: 
$\omega_{AB}=\omega_0+\delta\omega_0(t)$ and $\Omega=\Omega_0+\delta\Omega(t)$. The source of the noise 
in the considered system is due to the noise in the amplitude of external longitudinal and transverse magnetic fields. To obtain the master equation of two atoms 
interacting with fluctuating magnetic fields,  first $H(t)$ is transformed to the rotating frame of the transverse field with the standard transformation 
$M=exp(i\omega t\sum_{i=A,B}\sigma_z^i/2)$ with the aim to eliminate the sinusoidal time dependence of the transverse field. In this frame, 
the Hamiltonian is given as $H^{'}(t)=MH(t)M^{\dagger}-i M\dot{M}^{\dagger}=H_0^{'}+\delta H^{'}(t)$, where 
\begin{eqnarray}\label{hamro}
H_0^{'}&=&\frac{1}{2}[\Delta_{0}(\sigma_z^A+\sigma_z^B)+\Omega_{0}(\sigma_x^A+\sigma_x^B)],\nonumber\\
\delta H^{'}(t)&=&\frac{1}{2}[\delta\Delta (t)(\sigma_z^A+\sigma_z^B)+\delta\Omega (t)(\sigma_x^A+\sigma_x^B)],
\end{eqnarray}
and $\Delta_0=\omega_0-\omega$ is the detuning and $\delta\Delta (t)$ is its noisy component caused solely because of the noise in $\omega_{AB}$. 
The dynamics of the system density matrix, $\rho^{'}(t)$, in the rotating frame can be given by Liouville master equation:
 $\frac{\partial }{\partial t}\rho^{'}(t)=-i[H^{'}(t),\rho^{'}(t)]$. Carrying it into interaction picture changes this master
equation as $\frac{\partial}{\partial t}\rho^{'}_{I}(t)=-i[\delta H_I^{'}(t),\rho^{'}_{I}(t)]$ where 
$ \rho^{'}_{I}(t)=e^{iH_{0}^{'}t}\rho^{'}(t)e^{-iH_{0}^{'}t}$ and $\delta H^{'}_{I}(t)=e^{iH_{0}^{'}t}\delta H^{'}(t)e^{-iH_{0}^{'}t}$. 
The formal solution  of this equation can be given as 
\begin{eqnarray}\label{eq:formal}
\rho_{I}^{'}(t)&=&\rho_{I}^{'}(0)-i \int_0^t\,dt_1 [\delta H_{I}^{'}(t_1),\rho_{I}^{'}(0)]\nonumber\\
&+& (-i)^{2} \int_0^t\,dt_1\int_0^t\,dt_2 [\delta H_{I}^{'}(t_1),[\delta H_{I}^{'}(t_2),\rho_{I}^{'}(0)]]+\dots\nonumber\\
\end{eqnarray}
Since the formal solution in Eq.~(\ref{eq:formal}) has randomly fluctuating terms, we perform a stochastic average over the noise. 
After this process, the solution of the stochastic Liouville equation can be represented by the cumulant expansion technique introduced by Kubo~\cite{kubo}:
\begin{equation}
\label{eq:kubo}
\rho_{I}^{'}(t)=\exp_{p}\left(\sum_{n=1}^\infty K_n(t)\right)\rho_{I}^{'}(0),
\end{equation}
where $K_n(t)$ is the cumulant and the subscript $p$ indicates the partial-ordering prescription. We insert Eq.~(\ref{eq:kubo}) 
into stochastic Liouville equation and perform classical Gaussian stochastic Markov approximation; setting all 
cumulants, expect $K_{2}$, to be zero, then we obtain the master equation as:
\begin{eqnarray}\label{eq:cum}
\frac{\partial }{\partial t}\rho'_{I}(t)=\dot{K_2}\rho'_{I}(t)=-\int_0^t\langle[\delta H_{I}^{'}(t),[\delta H_{I}^{'}(t_1),\rho'_{I}(t)]]\rangle\,dt_1,\nonumber\\
\end{eqnarray}
where $\langle\dots\rangle$ represents stochastic average. Now we assume independent white-noise case: 
$\langle\delta h_i(t)\delta h_j(t')\rangle=\Gamma_{ h_i}\delta_{ij}\delta(t-t')$ ($h_{1}\equiv \Delta$, 
$h_{2}\equiv \Omega$) with zero mean $\langle\delta h_i(t)\rangle=0$. Here $\Gamma_{\Delta}$ and $\Gamma_{\Omega}$ 
are the (mean) noise strengths associated with the noise in the detuning and the Rabi frequency, respectively. 
White-noise approximation (also known as Markovian approximation) has no memory effects and is valid when the correlation 
time scale of the noise is much smaller than the evolution time scale of the system. To get a rigorous form of the master equation, 
we apply white-noise approximation to the master equation, Eq.~(\ref{eq:cum}), and transform back to the Shr\"{o}dinger picture. We finally get the 
master equation in the rotating frame as:
\begin{eqnarray}\label{mastereq}
\frac{\partial }{\partial t}\rho'(t)&=&-i[H_{0}',\rho'(t)]-\frac{\Gamma_{\Delta}}{4}[\sigma_{z}^A+\sigma_{z}^B,[\sigma_{z}^A+\sigma_{z}^B,\rho'(t)]]\nonumber\\
&-&\frac{\Gamma_{\Omega}}{4}[\sigma_{x}^A+\sigma_{x}^B,[\sigma_{x}^A+\sigma_{x}^B,\rho'(t)]].
\end{eqnarray}

It is worth noting that local unitary transformations do not change the eigenvalues of the system (as broadly speaking the correlation measures, 
such as quantum discord, entanglement, etc., do not change under local unitary transformations). Therefore, we will work in rotating frame and drop 
the superscript $"'"$ in the following.

We should emphasize here that in the absence of noise in detuning and Rabi frequency~(i.e., $\Gamma_{\Delta}=\Gamma_{\Omega}=0$), 
the master equation has only the unitary dynamics described by $H_0$ and its solution can be given as $\rho(t)=U^{\dagger}(t)\rho(0)U(t)$ 
where $U(t)=e^{-iH_0t}$ and $H_0$ is given in Eq.~(\ref{hamro}). Since $U(t)$ is both local and unitary in time, the correlations 
of a given state at any time under unitary evolution is exactly equal to the initial state correlations. That is, the unitary 
evolution alone is incapable of changing the correlation dynamics of a given input state. Moreover,
the master equation~(\ref{mastereq}) has no analytical solution. Therefore, we will put ourselves some legitimate limitations. 
In the following, our main concern will be the sole effect of the noise components in the master equation on quantum correlations, 
so, without loss of generality, we will set $\omega_0=\omega$ and $\Omega_0=0$ so that $i[\rho(t),H_0]=0$; no unitary evolution. 
Furthermore, we will consider X-structured density matrix defined by its elements $\rho_{12}=\rho_{13}=\rho_{24}=\rho_{34}=0$. 
Indeed, the dynamics under $\omega_0=\omega$ and $\Omega_0=0$ preserves the X-structure of the density matrix~\cite{xspre}. Under these restrictions, 
the dynamics can be described analytically, but these equations are quite cumbersome, so we will not display them here for brevity. On the other hand, the solution of the master equation
for a general family of states and $i[\rho(t),H_0]=0$ can also be obtained and is given in Ref.~\cite{daffer} in the operator-sum representation. Note that at the limiting case, $i[\rho(t),H_0]=0$ , the dynamics given by master equation~(\ref{mastereq}) can also be recognized as the dynamics of two qubits that interact with an environment possessing random signal noise in $x$ and $z$ directions or the dynamics of two spin 1/2 particle interacting with  fluctuating magnetic fields directed in $x$ and $z$ axes~(see Refs.~\cite{daffer,adkw} for details).

Now, we will briefly review the quantum correlation measures, entanglement and quantum discord, 
that will be considered in the present work. Entanglement of formation~(EoF) is
as a measure of entanglement which gives zero for separable states and one for maximally 
entangled (Bell) states~\cite{eofdef}. For two-qubit system, it is given as:
\begin{eqnarray}\label{eof}
EoF(\rho)=h\left[\frac{1+\sqrt{1-C^2(\rho)}}{2}\right],
\end{eqnarray}
where $h[x]=-x\log_2x-(1-x)\log_2(1-x)$ is the binary entropy and $C(\rho)=2\max\{0,|\rho_{14}|-\sqrt{\rho_{22}\rho_{33}},|\rho_{23}|-\sqrt{\rho_{11}\rho_{44}}\}$ 
is the concurrence for X-structured density matrix.

Quantum discord is defined as the difference between the quantum version of two classically equivalent definitions of 
mutual information~\cite{howz}. It is defined as the difference between total correlations, as quantified by mutual information, 
and classical correlations~(CC). Although, EoF and QD are equal to each other 
for pure states, the relation is complicated for mixed states; QD is believed to capture more general than entanglement type quantum correlations, since it can be non-zero for some mixed separable states, but we should stress that QD and EoF are different quantifiers of quantum correlations. Although, the definition of QD requires a complex extremization procedure, this can be done analytically for X-structured density matrix~\cite{maarga,wlnl}. Here we use the results given in Ref.~\cite{wlnl} where the calculation of 
QD is based on the positive-operator-valued measurements locally performed on the subsystem B. The QD and CC are given as~\cite{wlnl}
\begin{eqnarray}\label{qdcc}
QD(\rho)=\min\{Q_1,Q_2\},\quad CC(\rho)=\max\{CC_1,CC_2\},\nonumber\\
\end{eqnarray}
where $CC_j=h[\rho_{11}+\rho_{22}]-D_j$ and $Q_j=h[\rho_{11}+\rho_{33}]+\sum_{k=1}^4\lambda_k\log_2\lambda_k+D_j$
with $\lambda_k$ being the eigenvalues of $\rho$ and $h[x]$ is the binary entropy defined above. 
Here $D_1=h[\tau]$ where $\tau=\left(1+\sqrt{[1-2(\rho_{33}+\rho_{44})]^2+4(|\rho_{14}|+|\rho_{23}|)^2}\right)/2$ 
and $D_2=-\sum_{k=1}^4 \rho_{kk}\log_2\rho_{kk}-h[\rho_{11}+\rho_{33}]$. We should emphasize that in general, quantum discord is 
not a symmetric quantity, i.e., its value can depend the measurements performed on subsystem $A$ or $B$. However, in the present study, 
we will mainly consider the steady state correlations and for steady states we have $\rho_{22}^{SS}=\rho_{33}^{SS}$, so $S(\rho_A^{SS})=S(\rho_B^{SS})$, 
where $S(\rho)=-Tr(\rho\log_2\rho)$ is the von-Neumann entropy and $\rho_{A,B}=Tr_{B,A}\rho$; irrespective of whether the measurement is performed locally on subsystem $A$ or $B$.
\section{Results}
\subsection{Dynamics of correlations under global stochastic independent noises}\label{sec3}
In the following, we will analyze how the transverse ($\Gamma_{\Omega}$) and/or longitudinal ($\Gamma_{\Delta}$) 
noise components in the master equation can create quantum correlations between initially uncorrelated qubits and how these 
noises can affect preexisting quantum correlations initially encoded to the qubits. To do this, we will consider the product states 
of the form, $\left|gg\right\rangle$, $\left|ee\right\rangle$,  $\left|eg\right\rangle$ and $\left|ge\right\rangle$ as well as four types of Bell states,
 $\left|\Psi^{\pm}\right\rangle=1/\sqrt{2}(\left|ee\right\rangle\pm\left|gg\right\rangle)$ and
 $\left|\Phi^{\pm}\right\rangle=1/\sqrt{2}(\left|eg\right\rangle\pm\left|ge\right\rangle)$ as the initial states. 
  
First, we consider the sole effect of the detuning noise. For the system parameters considered in the present study~($\omega_0=\omega$, $\Omega_0=0$) and $\Gamma_{\Omega}=0$, this noise is effectively a Markovian dephasing process which acts globally on the qubits~\cite{jql1}. The time-dependent density matrix of the system under such a process, starting from X-shaped initial state, can be easily found as:
\begin{eqnarray}
\label{xgsmdp}
\rho(t)=\left (\begin{array}{cccc} \rho_{11}(0)  & 0 & 0  & \rho_{14}(0)e^{-4\Gamma_{\Delta}t} \\ 0  & \rho_{22}(0) & \rho_{23}(0)  & 0 \\ 0 
 & \rho_{32}(0) & \rho_{33}(0)  & 0 \\ \rho_{41}(0)e^{-4\Gamma_{\Delta}t}   & 0 & 0  & \rho_{44}(0) \end{array} \right) \ . \nonumber\\
\end{eqnarray}
The dynamics of entanglement, quantum discord and classical correlations in $\rho(t)$ of Eq.~(\ref{xgsmdp}) can be expressed analytically from Eqs.~(\ref{eof}) and~(\ref{qdcc}). The main results for detuning noise alone can be summarized as follows: {\bf (i)} For the $\left|\Psi^{\pm}\right\rangle$ initial Bell states, the entanglement as well as quantum discord decrease exponentially, while classical correlations remain as 1 for all times. {\bf (ii)} $\left|gg\right\rangle$, $\left|ee\right\rangle$, $\left|eg\right\rangle$ and $\left|ge\right\rangle$ separable and $\left|\Phi^{\pm}\right\rangle$ entangled initial states are prone to the effects of longitudinal noise, which is expected since such states form the decoherence free subspace~(DFS)~\cite{dmmo} of the considered dynamics. The detuning noise alone leads to a simple dynamics which does not have any interesting behavior for the quantum correlations. So, in the following, we will mainly consider the effect of transverse noise alone~(i.e., $\Gamma_{\Delta}=0,\Gamma_{\Omega}\neq 0$) and both transverse and longitudinal noises~(i.e., $\Gamma_{\Delta}\neq 0,\Gamma_{\Omega}\neq 0$).

It is found that neither transverse noise~($\Gamma_{\Omega}$) alone nor longitudinal and transverse noises acting together can create entanglement from initially product states. On the other hand, both noise combinations lead to steady states with relatively high quantum discord. Below, we present and discuss such states.
\begin{figure}[!hbt]\centering
{\scalebox{0.41}{\includegraphics{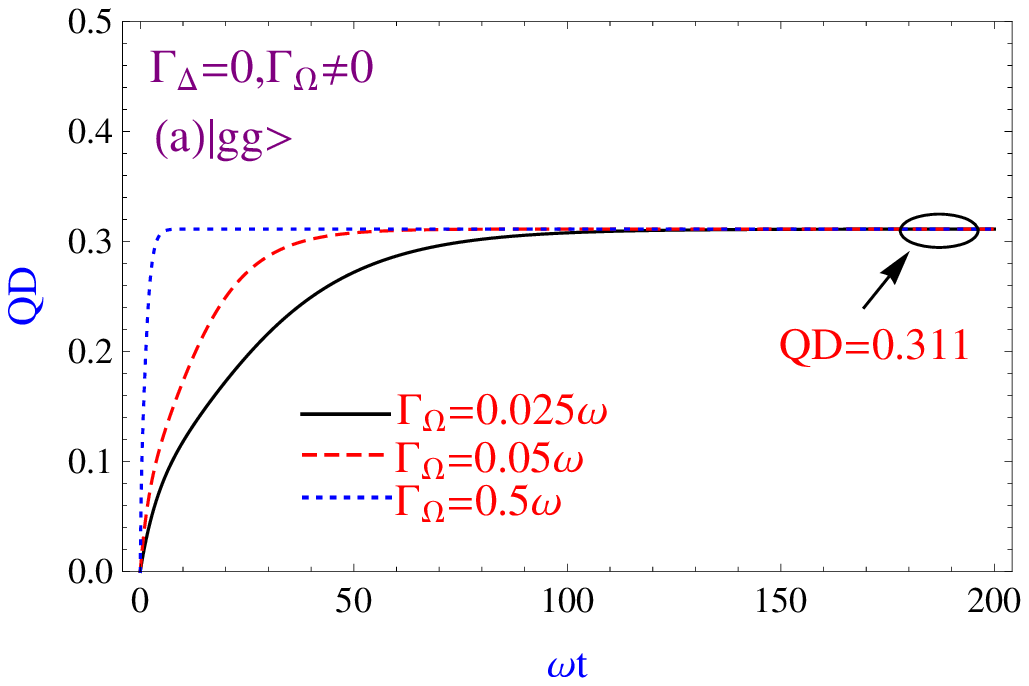}}}
{\scalebox{0.41}{\includegraphics{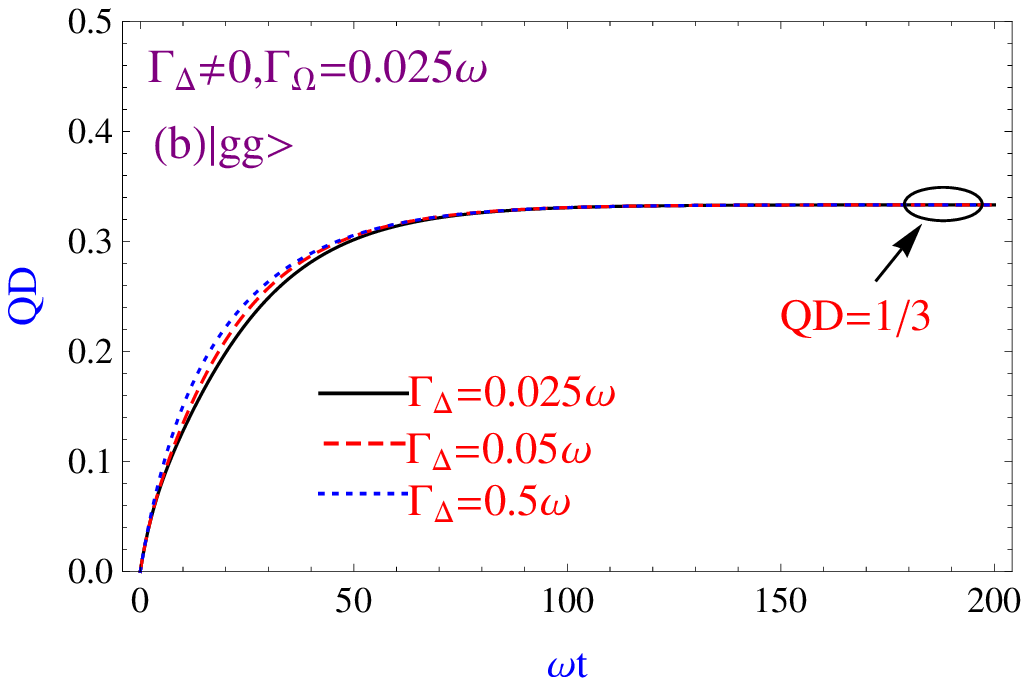}}}

{\scalebox{0.41}{\includegraphics{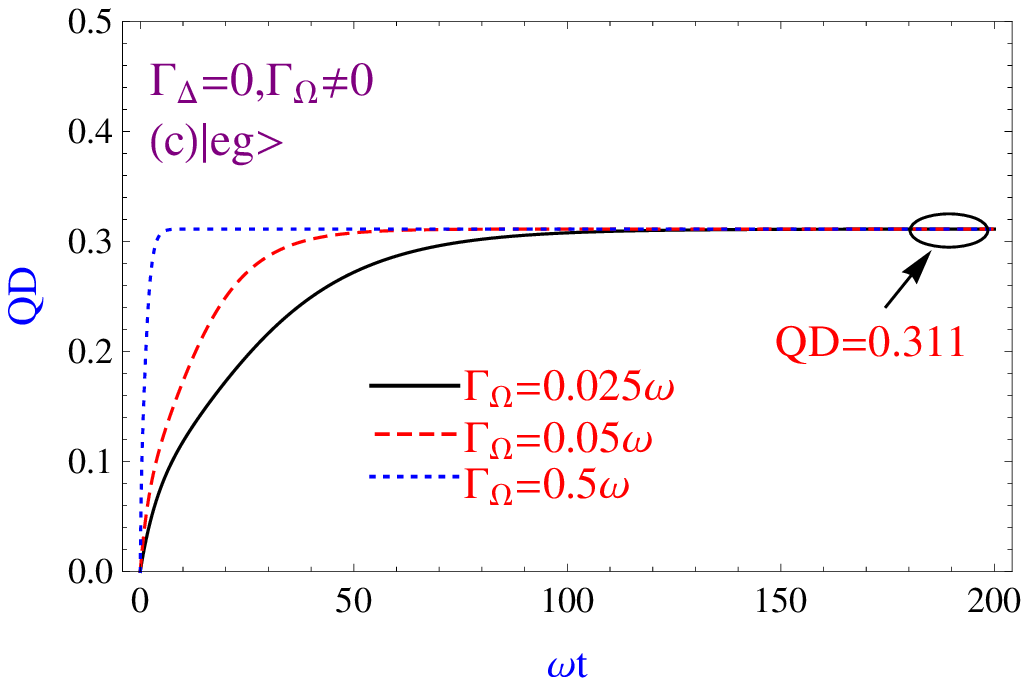}}}
{\scalebox{0.41}{\includegraphics{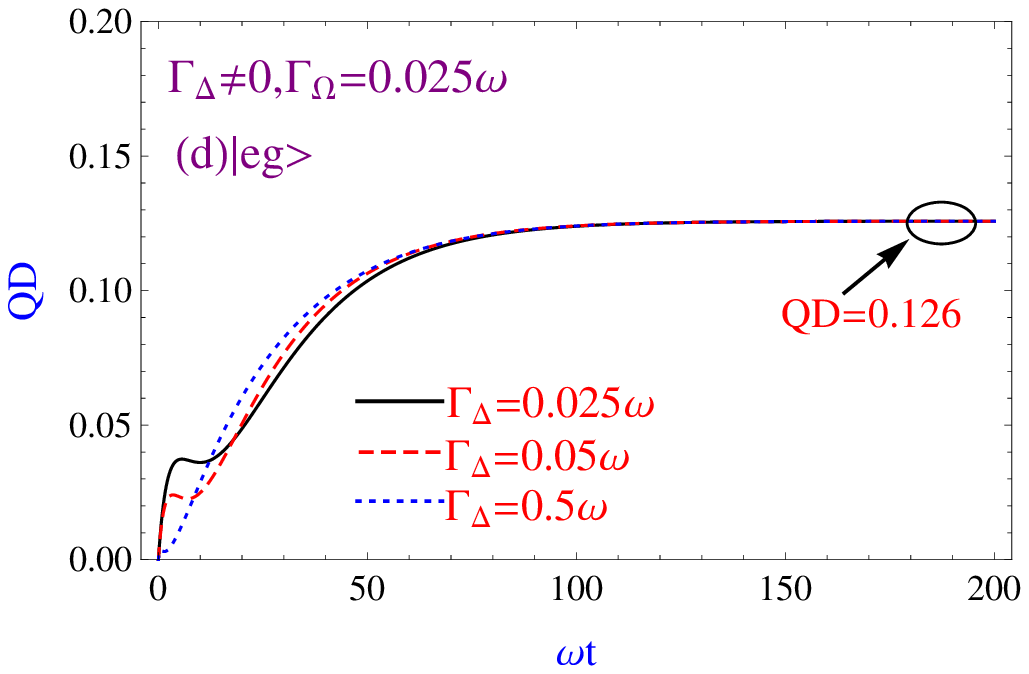}}}
\caption{Quantum discord versus $\omega t$ for $\left|gg\right\rangle$~(or  $\left|ee\right\rangle$)~(a) and~(b) 
and $\left|eg\right\rangle$~(or  $\left|ge\right\rangle$)~(c) and~(d) initial states for the cases $\Gamma_{\Delta}=0,\Gamma_{\Omega}\neq 0$~(left column) 
and $\Gamma_{\Delta}\neq 0,\Gamma_{\Omega}\neq 0$~(right column). Note that for each considered case, entanglement is not induced, so they are not plotted here.}
\end{figure}

In Fig.~1, we plot quantum discord versus dimensionless time, $\omega t$, for $\left|gg\right\rangle$~(Figs.~1(a) and~1(b)) 
and $\left|eg\right\rangle$~(Figs.~1(c) and~1(d)) initial states for different noise strengths with the aim to analyze initial 
state dependence of the noise-induced quantum correlations. One should note that the results obtained for $\left|gg\right\rangle$ and 
$\left|eg\right\rangle$ are the same as that for $\left|ee\right\rangle$ and $\left|ge\right\rangle$ initial states, respectively. 
We have observed that for every 
possible values of $\Gamma_{\Delta}$ and $\Gamma_{\Omega}$ and for the considered initial states, $EoF(t)=0$ for $t\geq 0$; no entanglement is induced. Contrary to EoF, the quantum correlations beyond entanglement, such as quantum discord can be easily created 
by the stochastic noises~\cite{facca} which also drive the system into steady states with extremely high quantum correlations; 
for $\left|gg\right\rangle$ initial state $QD=0.311$ for $\Gamma_{\Delta}=0,\Gamma_{\Omega}\neq 0$ case~(Fig.~1(a)) and $QD=1/3$ 
for $\Gamma_{\Delta}\neq 0,\Gamma_{\Omega}\neq 0$ case~(Fig.~1(b)), while for $\left|ge\right\rangle$ initial state $QD=0.311$ for 
$\Gamma_{\Delta}=0,\Gamma_{\Omega}\neq 0$ case~(Fig.~1(c)) and $QD=0.126$ for $\Gamma_{\Delta}\neq 0,\Gamma_{\Omega}\neq 0$ case~(Fig.~1(d)). 
Moreover, for each of the initial states considered here, the steady state value of QD is independent of the noise strength which only determines how fast the steady state is reached for each noise combination. Recently, Al-Qasimi and James~\cite{aqdj}, and independently Galve {\it et al.,}~\cite{ggz} 
showed that the maximum value of QD for two separable qubits is 1/3. Remarkably, the competition between the two independent noises in the system
 not only enhances the steady state QD, but it saturates the maximum possible QD for $\left|gg\right\rangle$ initial state~(see Fig.~1(b)). The steady state density matrix for
 two-active-noise case for $\left|gg\right\rangle$ initial state can be given as $\rho^{SS}=1/3\left(\left|\Phi^{+}\right\rangle\left\langle \Phi^{+}\right|+
\left|ee\right\rangle\left\langle ee\right|+\left|gg\right\rangle\left\langle gg\right|\right)$ which is, indeed, the 
local transform of the maximally discordant mixed state, $\rho=1/3\left(\left|\Psi^{+}\right\rangle\left\langle \Psi^{+}\right|+
\left|eg\right\rangle\left\langle eg\right|+\left|ge\right\rangle\left\langle ge\right|\right)$ for unentangled qubits  given in Ref.~\cite{ggz}. The dynamics of QD for the set of initially separable states considered here are identical in the case of the noise in the Rabi frequency alone~(Figs.~1(a) and~1(c)), while two simultaneous noises~($\Gamma_{\Delta}\neq 0,\Gamma_{\Omega}\neq 0$) effect the dynamics of $\left|gg\right\rangle$ and $\left|ee\right\rangle$ states different than $\left|eg\right\rangle$ and $\left|ge\right\rangle$ initial states. Introducing non-zero $\Gamma_{\Delta}$ noise above $\Gamma_{\Omega}$ increases~(decreases) the steady state QD of  $\left|gg\right\rangle$ and $\left|ee\right\rangle$~($\left|eg\right\rangle$ and $\left|ge\right\rangle$) initial states from 0.311~(0.311) to 1/3~(0.126).

Now, we consider the dynamics of quantum correlations under transverse alone and transverse plus longitudinal noises for the initially maximally quantum-correlated states. 
For that purpose, we take the Bell states in the form 
$\left|\Psi^{\pm}\right\rangle=1/\sqrt{2}(\left|ee\right\rangle\pm\left|gg\right\rangle)$ and
 $\left|\Phi^{\pm}\right\rangle=1/\sqrt{2}(\left|eg\right\rangle\pm\left|ge\right\rangle)$ as the initial states. Before starting our qualitative analysis, 
we should stress that $\left|\Phi^{-}\right\rangle$ is the decoherence free subspace of the overall dynamics considered in this Letter. This is expected since the full Hamiltonian 
given by Eq.~(\ref{hamiltonian}) provides a rotation and the state $\left|\Phi^{-}\right\rangle$ remains unchanged by this rotation.
\begin{figure}[!hbt]\centering
{\scalebox{0.41}{\includegraphics{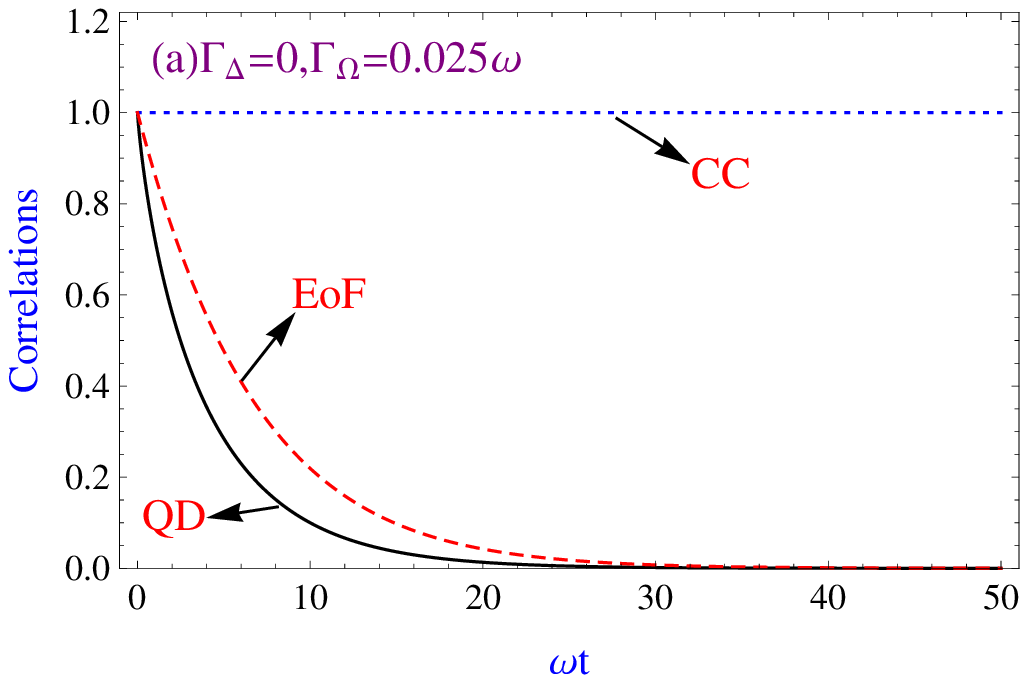}}}
{\scalebox{0.41}{\includegraphics{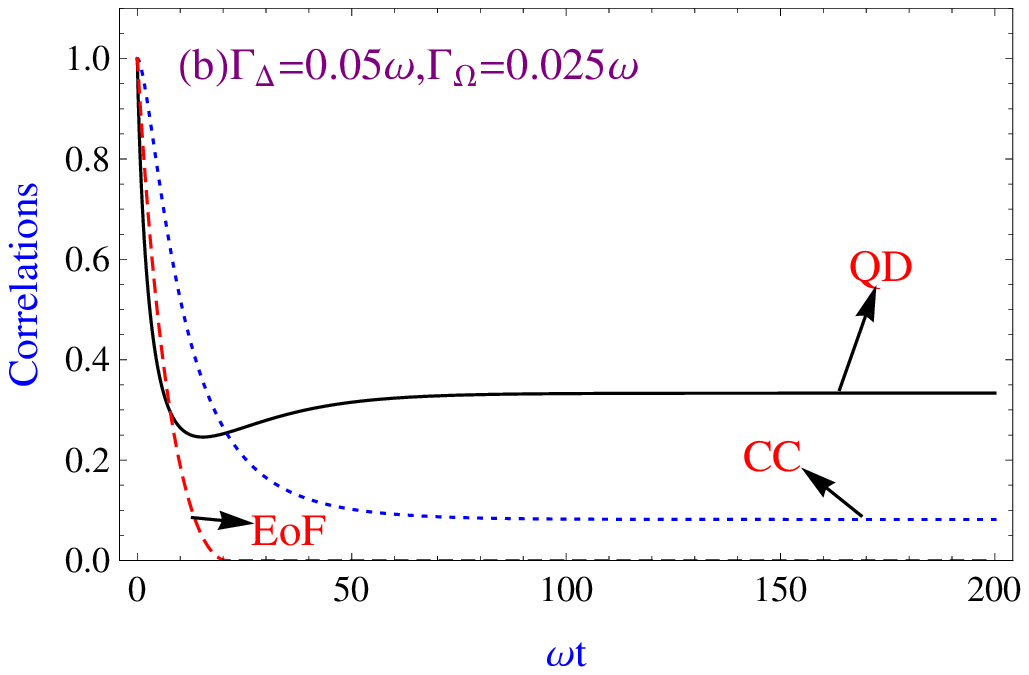}}}
\caption{QD~(black), EoF~(red, dashed), CC~(blue, dotted) versus $\omega t$ for $\left|\Phi^{+}\right\rangle$ initial state for
 the cases (a)$\Gamma_{\Delta}=0,\Gamma_{\Omega}\neq 0$ and (b)$\Gamma_{\Delta}\neq 0, \Gamma_{\Omega}\neq 0$.}
\end{figure}

In Fig.~2, we display the dynamics of QD, EoF and CC versus the dimensionless time $\omega t$ under $\Gamma_{\Omega}$ noise~(Fig.~2(a)) and under collective, 
$\Gamma_{\Omega}$ and $\Gamma_{\Delta}$, noises~(Fig.~2(b)) for the $\left|\Phi^{+}\right\rangle$ initial state. Contrary to initial product states case, 
the $\Gamma_{\Omega}$ noise is highly detrimental to initial quantum correlations present in the system as quantified by EoF and QD which decay exponentially to zero. Peculiarly, CC
is unaffected by this noise in the overall dynamics as indicated by blue-dotted line in Fig.~2(a). In fact, such dynamical behavior can immediately suggest
an operational way of computing quantum discord without any extremization procedure~\cite{mcsv}. QD for such cases can be given as the difference between the 
evolved state mutual information and the completely decohered state mutual information. On the other hand, when two independent noises are considered to be non-zero,
 EoF dies quickly, while QD reaches a steady value~(Fig.~2(b)). It is safe to suggest that more independent noises in the system, although highly detrimental to entanglement,
 can be highly constructive for quantum correlations beyond entanglement, such as quantum discord. 
It is interesting to note that both EoF and QD decay monotonically with time until entanglement completely vanishes. After the time where ESD happens, 
QD starts to increase and reaches a steady state which is, in fact, the maximally discordant mixed state for separable states given by 
$\rho^{SS}=1/3\left(\left|\Phi^{+}\right\rangle\left\langle \Phi^{+}\right|+\left|ee\right\rangle\left\langle ee\right|+\left|gg\right\rangle\left\langle gg\right|\right)$. 
Similar to EoF, more independent noises in the system are also detrimental  to classical correlations for this particular initial state 
as can be seen from the comparison of Figs.~2(a) and~2(b). One should note that we have plotted only one fixed $\Gamma_{\Omega}$ and/or $\Gamma_{\Delta}$. 
The noise strength in each case, in fact, only determines the time in which the correlations attain their steady values rather than the magnitudes.

We have also considered the other initial Bell states in the form $\left|\Psi^{\pm}\right\rangle=1/\sqrt{2}(\left|ee\right\rangle\pm\left|gg\right\rangle)$. Under $\Gamma_{\Omega}$-type noise, the dynamics of correlations for $\left|\Psi^+\right\rangle$ 
state is exactly the same as that of $\left|\Phi^+\right\rangle$ as depicted in Fig.~2(a), while for $\left|\Psi^-\right\rangle$ state, the correlations 
are found to be unaffected by $\Gamma_{\Omega}$ noise;  $\left|\Psi^-\right\rangle$ is also the DFS of the dynamics under the conditions 
$\Delta_0=0$, $\Omega_0=0$ and $\Gamma_{\Delta}=0$. These are expected, since the Hamiltonian given in Eq.~(\ref{hamiltonian}) 
provides rotations around some axes by a given angle, the Bell states can be connected to each other by these local transformations 
or not to be affected. On the other hand, the steady state properties of $\left|\Psi^{\pm}\right\rangle$ state for 
$\Gamma_{\Delta}\neq 0$, $\Gamma_{\Omega}\neq 0$ are found to be the same as that of $\left|\Phi^+\right\rangle$ initial state case as depicted in Fig.~2(b), but the dynamical behaviors of the correlations are qualitatively different; we mean that the times where ESD 
and saturation happen and the minimum that QD can attain in the time evolution can become different for the initial states 
$\left|\Psi^{\pm}\right\rangle$ and $\left|\Phi^+\right\rangle$. Therefore, these are not plotted here.
\newcommand{\ra}[1]{\renewcommand{\arraystretch}{#1}}
\begin{table*}\centering

\small
\begin{tabular}{@{}rrrrrrcrrrrr@{}}\toprule
\hline
Initial state & \multicolumn{5}{c}{$\Gamma_{\Delta}=0$, $\Gamma_{\Omega}\neq 0$} & \phantom{abc}& \multicolumn{5}{c}{$\Gamma_{\Delta}\neq 0$, $\Gamma_{\Omega}\neq 0$}  \\
\cmidrule{2-6} \cmidrule{8-11} 
& EoF & QD & GMQD & CC & S$_L$ && EoF & QD & GMQD & CC & S$_L$  \\ 
\hline\hline
  
$\left|gg\right\rangle$,$\left|ee\right\rangle$ & 0 & 0.311 & 0.0625 & 0.189 & 0.833 && 0 & 1/3 &  0.0556 & 0.0817 & $8/9$ \\

$\left|eg\right\rangle$,$\left|ge\right\rangle$ & 0 & 0.311 & 0.0625 & 0.189 & 0.833 && 0 & 0.126 & 0.0556 & 0.0817 & $8/9$  \\

$\left|\Phi^+\right\rangle$ & 0 & 0 & 0 & 1 & $6/9$ && 0 & 1/3 & 0.0556 & 0.0817 & $8/9$  \\

$\left|\Phi^-\right\rangle$ & 1 & 1 & 0.5 & 1 & 0 && 1 & 1 & 0.5 & 1 & 0 \\

$\left|\Psi^+\right\rangle$ & 0 & 0 & 0 & 1 & 6/9 && 0 & 1/3 & 0.0556 & 0.0817 & 8/9 \\

$\left|\Psi^-\right\rangle$ & 1 & 1 & 0.5 & 1 & 0 && 0 & 1/3 & 0.0556 & 0.0817 & 8/9 \\

\hline

\end{tabular}
\caption{The steady state correlations~(EoF, QD, GMQD, CC) and the linear entropy, $S_L$, for the initially product and Bell states under the conditions $\Delta_0=0$ and $\Omega_0=0$ for the cases  $\Gamma_{\Delta}=0$, $\Gamma_{\Omega}\neq 0$ and $\Gamma_{\Delta}\neq 0$,
 $\Gamma_{\Omega}\neq 0$. Note that in the absence of entanglement~($EoF=0$), $QD=1/3$ is the maximum possible value that can attain for separable states.}
\end{table*}

Now, we collect the steady-state values of EoF, QD, CC as well as the geometric measure of quantum discord~(GMQD)~\cite{gmqd} 
introduced by Dakic, Vedral and Brukner in Ref.~\cite{dvb} and the linear entropy, $S_L=\frac{4}{3}[1-Tr(\rho^2)]$ for each considered initial state and noise combinations and present them in Table~1. The most important finding in Table~1 is the saturation of maximal QD~($QD=1/3$) for separable qubits under the effect of both transverse and longitudinal noises for $\left|gg\right\rangle$, $\left|ee\right\rangle$,$\left|\Phi^+\right\rangle$ and $\left|\Psi^{\pm}\right\rangle$ initial states. In fact, the enhancement of the steady state QD in the absence of entanglement with two independent noises for $\left|gg\right\rangle$, $\left|ee\right\rangle$,$\left|\Phi^+\right\rangle$ and $\left|\Psi^{+}\right\rangle$ initial states can be explained by considering discord-linear entropy relation as was done in Ref.~\cite{aqdj}. It is well known that $S_L$ is a measure of randomness, so it is expected that the quantumness 
and randomness exhibit an inverse relationship. Indeed, in general, QD decreases as linear entropy increases, or vice versa. On the other hand, there exist 
some unentangled mixed states in which QD and randomness show peculiarly linear dependence. As can be seen from the tabulated  numerical values in Table~1, for the case $\Gamma_{\Delta}=0,\Gamma_{\Omega}\neq 0$, $S_L=0.833$ for $\left|gg\right\rangle$ and $\left|ee\right\rangle$ with steady state $QD=0.311$, and $S_L=6/9$ for $\left|\Psi^+\right\rangle$ and $\left|\Phi^+\right\rangle$ with steady state $QD=0$, while $S_L=8/9$ for such initial states with steady state $QD=1/3$. Indeed, two independent noises in the system can increase both the randomness and the relevant quantumness measure, 
quantum discord. Moreover, we should stress that $QD=1/3$ is the maximum possible value that can be reached as an increasing 
function of randomness in the absence of entanglement~\cite{aqdj}. Contrary to QD, considering the effect of two independent noises in the system on CC shows that the steady state CC~(also the total 
correlations given by the sum of discord and classical correlations) decrease as the randomness increases for $\left|gg\right\rangle$, $\left|ee\right\rangle$,$\left|\Phi^+\right\rangle$ and $\left|\Psi^+\right\rangle$ initial states. It is interesting to note that GMQD disagree with the enhancement of QD  with randomness for $\left|gg\right\rangle$, $\left|ee\right\rangle$ initial states. On the other hand, 
for $\left|eg\right\rangle$ and $\left|ge\right\rangle$ initial states, all types of steady state correlations, including QD, decrease when introducing non-zero $\Gamma_{\Delta}$ noise above $\Gamma_{\Omega}$. Note that for $\Gamma_{\Delta}\neq 0,\Gamma_{\Omega}\neq 0$ case, the steady state properties, 
such as CC, GMQD and $S_L$ have exactly the same values for the set of initial states $\left|gg\right\rangle$, $\left|ee\right\rangle$,$\left|\Phi^+\right\rangle$ and $\left|\Psi^{\pm}\right\rangle$  and $\left|eg\right\rangle$, while QD is quite different. 
The steady state for the set of initial states is $\rho^{SS}=1/3\left(\left|\Phi^{+}\right\rangle\left\langle \Phi^{+}\right|+\left|ee\right\rangle\left\langle ee\right|+\left|gg\right\rangle\left\langle gg\right|\right)$, 
while for $\left|eg\right\rangle$, $\rho^{SS}(\left|eg\right\rangle)=\frac{1-\epsilon}{4}I_4+\epsilon\left|\Phi^-\right\rangle\left\langle\Phi^-\right|$ with $\epsilon=1/3$. 
Although these two steady states have the same single qubit von-Neumann entropies, $S(\rho^{SS}(\left|eg\right\rangle))>S(\rho^{SS})$. Thus the 
reduction of steady state QD
with two independent active noise for $\left|eg\right\rangle$ and $\left|ge\right\rangle$  initial states compared to the enhancement of steady QD for $\left|gg\right\rangle$ and $\left|ee\right\rangle$ initial states as shown in Fig.~1 can be understood as more loss of total correlations in the steady states as measured by mutual information.
\subsection{Classical noise induced maximally discordant mixed separable steady states}\label{sec4}
In the previous section, it was shown that the competition between two independent noises gives rise to not only a relatively high QD, but indeed it maximizes the QD
in the absence of entanglement for some initially product and Bell states. On the other hand, the individual noises, $\Gamma_{\Delta}$ and $\Gamma_{\Omega}$, were found to 
be insufficient to maximize the steady-state QD. It is natural to ask: do the individual noises in the system maximize the steady state QD for separable states ? To answer 
this question, in this section we will determine the conditions for which the steady state QD is maximized in the absence of entanglement by the $\Gamma_{\Delta}$ and/or $\Gamma_{\Omega}$ type noises.

We first consider the ability of the noise in the detuning to maximize the QD for separable states. The steady states for a system initially in a X-structured 
density matrix for the case $\Gamma_{\Delta}\neq 0$,$\Gamma_{\Omega}=0$ can be easily found as: $\rho_{ii}^{SS}=\rho_{ii}(0)$~($i=1,2,3,4)$, $\rho_{23}^{SS}=\rho_{23}(0)$ and 
$\rho_{14}^{SS}=0$. Due to the simple structure of the steady states, we can immediately suggest a class of initial states as $\rho(0)
=1/3\left(\left|\Phi^{\pm}\right\rangle\left\langle \Phi^{\pm}\right|+\left|ee\right\rangle\left\langle ee\right|+\left|gg\right\rangle\left\langle gg\right|\right)
+c\left|ee\right\rangle\left\langle gg\right|+c^*\left|gg\right\rangle\left\langle ee\right|$ where $\Gamma_{\Delta}$-type noise unambiguously maximize the steady 
state QD for separable states for every possible values of $c$ obeying  $0<|c|\leq 1/3$ that can be easily checked from the time-dependent density matrix, Eq.~(\ref{xgsmdp}).
\begin{figure}[!hbt]\centering
{\scalebox{0.55}{\includegraphics{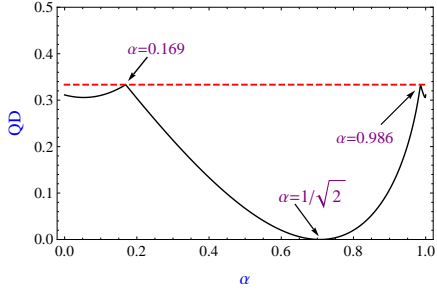}}}
\caption{$\alpha$  dependence of the steady state QD for $\left|\Phi_{\alpha}^+\right\rangle$ or $\left|\Psi_{\alpha}^+\right\rangle$ initial state under $\Gamma_{\Omega}$ noise. 
The red-dashed line  shows the maximum value of QD~($QD=1/3$) that can attain for separable states. Note that 
for every possible values of $\alpha$, the steady state entanglement is zero, so it is not plotted here, and the results for $\left|\Phi_{\alpha}^+\right\rangle$ and $\left|\Psi_{\alpha}^+\right\rangle$ initial states are the same.}
\end{figure}

On the other hand, for the case  $\Gamma_{\Delta}=0$, $\Gamma_{\Omega}\neq 0$, the steady states are highly initial state dependent and can be given as
\begin{widetext}
\begin{eqnarray}\label{case2ss}
\rho_{11}^{SS}&=&\rho_{44}^{SS}=\frac{1}{8} \left(3 \rho _{11}(0)-\rho _{14}(0)+\rho _{22}(0)+\rho _{23}(0)+\rho _{32}(0)+\rho _{33}(0)-\rho _{41}(0)+3 \rho _{44}(0)\right),\nonumber\\
\rho_{22}^{SS}&=&\rho_{33}^{SS}=\frac{1}{8} \left(\rho _{11}(0)+\rho _{14}(0)+3 \rho _{22}(0)-\rho _{23}(0)-\rho _{32}(0)+3 \rho _{33}(0)+\rho _{41}(0)+\rho _{44}(0)\right),\nonumber\\
\rho_{14}^{SS}&=&\frac{1}{8} \left(-\rho _{11}(0)+3 \rho _{14}(0)+\rho _{22}(0)+\rho _{23}(0)+\rho _{32}(0)+\rho _{33}(0)+3 \rho _{41}(0)-\rho _{44}(0)\right),\nonumber\\
\rho_{23}^{SS}&=&\frac{1}{8} \left(\rho _{11}(0)+\rho _{14}(0)-\rho _{22}(0)+3 \rho _{23}(0)+3 \rho _{32}(0)-\rho _{33}(0)+\rho _{41}(0)+\rho _{44}(0)\right).
\end{eqnarray}
\end{widetext}
We have searched for a general class of initial states by using several class of X-structured states, including Werner-like states and the other class of states given in Ref.~\cite{aqdj}
for which the steady state QD is maximized under $\Gamma_{\Omega}$ noise, but due to the quite involved initial state dependence 
of the steady states~(Eq.~(\ref{case2ss})), it seems to be not an easy task to determine  a general class of initial states without violating the density matrix properties. Nevertheless, we can use Bell-like initial states in the form as 
$\left|\Phi_{\alpha}^+\right\rangle=\alpha\left|eg\right\rangle+\sqrt{1-\alpha^2}\left|ge\right\rangle$ and 
$\left|\Psi_{\alpha}^+\right\rangle=\alpha\left|ee\right\rangle+\sqrt{1-\alpha^2}\left|gg\right\rangle$. Here $\alpha$~$(0\leq\alpha\leq 1)$ is called the degree of correlations since EoF
and QD monotonously increases from zero to 1 for $\alpha$ values from 0 (or 1) to $1/\sqrt{2}$~\cite{jql1}. In Fig.~3, we display the $\alpha$-dependence of the steady state QD for  
$\left|\Phi_{\alpha}^+\right\rangle$ and $\left|\Psi_{\alpha}^+\right\rangle$ initial states. In fact, the $\alpha$-dependence of steady state QD for these class of initial states 
are the same and the steady states contain no EoF. On the other hand, these class of initial states can immediately 
disclose that the steady state QD can also be maximized at $\alpha=0.169$ or $\alpha=0.986$ for separable states under $\Gamma_{\Omega}$-type noise. 
One should note that the initial and final 
state QD show peculiarly inverse $\alpha$-dependence, especially between $0.169<\alpha<0.986$; QD reaches high steady state value 
for $\alpha$-values where initial state QD is quite small or zero, while it becomes low and even zero~($\alpha=1/\sqrt{2}$) 
for $\alpha$-values where initial state QD is high and maximum~($\alpha=1/\sqrt{2}$).

As a final demonstration, we consider the steady states under the collective noise $\Gamma_{\Delta}\neq 0$, $\Gamma_{\Omega}\neq 0$, which are:
\begin{widetext}
\begin{eqnarray}\label{case3ss}
\rho_{11}^{SS}&=&\rho_{44}^{SS}=\frac{1}{6} \left(2 \rho _{11}(0)+\rho _{22}(0)+\rho _{23}(0)+\rho _{32}(0)+\rho _{33}(0)+2 \rho _{44}(0)\right),\nonumber\\
\rho_{22}^{SS}&=&\rho_{33}^{SS}=\frac{1}{6} \left(\rho _{11}(0)+2 \rho _{22}(0)-\rho _{23}(0)-\rho _{32}(0)+2 \rho _{33}(0)+\rho _{44}(0)\right),\nonumber\\
\rho_{14}^{SS}&=&0,\quad \rho_{23}^{SS}=\frac{1}{6} \left(\rho _{11}(0)-\rho _{22}(0)+2 \rho _{23}(0)+2 \rho _{32}(0)-\rho _{33}(0)+\rho _{44}(0)\right).\nonumber\\
\end{eqnarray}
\end{widetext}
In fact, due to the symmetric structure of the populations and coherences in steady states, we can suggest several class of initial states, such as Bell-like states in the form 
$\left|\Psi_{\alpha}^{\pm}\right\rangle=\alpha\left|ee\right\rangle\pm\sqrt{1-\alpha^2}\left|gg\right\rangle$ where the collective noise maximize the steady state QD in the absence of entanglement for every possible values of $\alpha$ between 0 and 1. Analyzing  Fig.~3 shows that Bell like initial states, $\left|\Psi_{\alpha}^+\right\rangle$, 
with $\alpha=0.169$ or $\alpha=0.986$ have maximally discordant mixed separable state under both the individual, $\Gamma_{\Omega}$, and the collective independent noises, 
$\Gamma_{\Delta}$,$\Gamma_{\Omega}$. Since $\rho_{14}^{SS}=0$ as shown in Eq.~(\ref{case3ss}), we are also able to suggest a more general class of initial states that includes Bell states, 
the so-called $\beta$-states in the form $\rho_{\beta}(0)=\beta\left|\Psi^+\right\rangle\left\langle\Psi^+\right|+(1-\beta)\left|\Phi^+\right\rangle\left\langle\Phi^+\right|$ 
where the steady state QD is maximal ($QD=1/3$)  in the absence of entanglement for every $\beta$ values between $0\leq\beta\leq 1$. $\beta$-states are known to form the lower 
bound for quantum discord for a given entanglement of formation~\cite{aqdj}. Intriguingly, they provide the upper bound of QD for EoF=0 under 
$\Gamma_{\Delta}\neq 0$, $\Gamma_{\Omega}\neq 0$ case.

Noise assisted creation of quantum correlations are sometimes attributed to back-action of the environment and/or 
memory effects~\cite{cnp}. On the other hand, the environment modeled as noisy external magnetic field in the present 
study is both classical and Markovian. Classical environments are not able to store quantum correlations, and the 
considered noises in the present problem is Markovian, so there are no memory effects. Because of these, there would 
be no back-action on the system via noisy magnetic fields here. A natural question that arise in the present case is 
the source of high steady state QD achieved even from initial product states. To answer this question, we analyze a 
problem which is similar to the one outlined above, but the noisy magnetic fields at each qubit act locally, and compare the 
steady states of this dynamics with that of the global magnetic field case. The Hamiltonian with the local fields can be given as: 
$H(t)=1/2\sum_{i=A,B}\{\delta\Delta_i(t)\sigma_z^i+\delta\Omega_i(t)\sigma_x^i\}$, with non-zero average noise strengths 
$\langle\delta\Delta_{A(B)}(t)\delta\Delta_{A(B)}(t')\rangle=\Gamma_{\Delta}^{A(B)}\delta(t-t')$ and  
$\langle\delta\Omega_{A(B)}(t)\delta\Omega_{A(B)}(t')\rangle=\Gamma_{\Omega}^{A(B)}\delta(t-t')$. 
Substituting the considered Hamiltonian into Liouville master equation and applying the considered stochastic 
averages yield, 
$\dot{\rho}=-\Gamma_{\Delta}/4\sum_{i=A,B}[\sigma_z^i,[\sigma_z^i,\rho]]-\Gamma_{\Omega}/4\sum_{i=A,B}[\sigma_x^i,[\sigma_x^i,\rho]]$ 
where $\Gamma_{\Delta}^A=\Gamma_{\Delta}^B=\Gamma_{\Delta}$ and  $\Gamma_{\Omega}^A=\Gamma_{\Omega}^B=\Gamma_{\Omega}$. 
It is straightforward to show that for an initially X-structured density matrix, the steady state density matrices would 
not contain coherence components under the individual, $\Gamma_{\Delta}$, and the collective, $\Gamma_{\Delta},\Gamma_{\Omega}$, noises. 
This means that the steady states contain no QD. On the other hand, under $\Gamma_{\Omega}$-type noise, the steady state populations would 
be equally weighted~(i.e., 1/4), while the coherences are $\rho_{23}^{SS}=\rho_{14}^{SS}=1/4(\rho_{14}(0)+\rho_{23}(0)+h.c.)$. 
In fact, this steady state is fully classically correlated since it is diagonal in the basis resulting from the tensor product of two 
local orthogonal basis, $\{\left|\pm_A\right\rangle\}$,$\{\left|\pm_B\right\rangle\}$, where $\left|\pm\right\rangle=1/\sqrt{2}(\left|e\right\rangle\pm\left|g\right\rangle)$. 
As a result, under local transverse and/or longitudinal memoryless classical noises, the steady states are fully quantum uncorrelated as quantified by QD and EoF. 
Mathematically speaking, the cross average noise strength terms, $\langle\delta\Delta_A(t)\delta\Delta_B(t')\rangle$ and $\langle\delta\Omega_A(t)\delta\Omega_B(t')\rangle$, 
that are considered to be non-zero for the global field case, are responsible for the steady states with high QD. This effect is sometimes regarded as the effect of 
common environment mediated indirect qubit-qubit interaction~\cite{cnp,mmpsg}.
\section{Conclusions}\label{sec5}
We have investigated the dynamics of quantum~(such as entanglement and quantum discord) and classical correlations as well as the steady state properties~(such as geometric measure of quantum discord and linear entropy)  for two qubits subject to longitudinal and/or transverse noisy magnetic fields.  We have shown that starting from different initial product or correlated states, each of the considered noises alone or both of them acting together can lead to steady states which can carry the maximum possible quantum discord for a separable state. Increase of quantum correlations under noisy conditions is sometimes attributed to the back-action of the environment or the memory effects in the environment-system interaction. In the model studied here, there is no back-action or the memory. To better understand the source of the steady state QD creation in the present model, we have also considered a similar model in which noise acts on the individual qubits locally and found that steady states contain no quantum correlations for this case. Therefore, the global nature of the noise is found to be responsible for the saturation of the maximally discordant mixed separable steady states.
Moreover, two independent noises in the system are found to enhance both the steady state randomness and the relevant quantumness measure, the so-called quantum discord, in the absence of entanglement. On the other hand, the geometrically defined version of QD~(GMQD) is found to disagree with the rise of steady state QD with randomness for some initial states.

One should note that the model system given by Hamiltonian, Eq.~(\ref{hamiltonian}), is an important model and the prototypical example of nuclear magnetic resonance which was also implemented recently in the observation of Berry's phase in a solid-state qubit~\cite{model2}. It would be very desirable to investigate the role of $\Delta_0$ and $\Omega_0$, which are neglected in the present Letter, together with their noisy components on quantum correlations for external fields acting locally or globally on qubits which is left for future investigations.


\begin{thebibliography}{99}
\bibitem{gmqd} The geometric measure of quantum discord measures the quantum correlations through the minimum Hilbert-Schmidt distance between the given state and zero discord state, in the sense of the fact that almost all quantum states have non-vanishing quantum discord. In our calculations, we have used the simplified analytic form of GQMD for two-qubit case provided in Ref.~\cite{dvb}.
\end{thebibliography}
\end{document}